\documentstyle[12pt]{article}

\newcommand{\mtx}[2]{\left(\begin{array}{#1}#2\end{array}\right)}

\begin{document}

\begin{center}

\bigskip
{\Large Parallel transport in an entangled ring}\\

\bigskip

William K.~Wootters\\

\bigskip

{\small{\sl

Department of Physics, Williams College, Williamstown, 
MA 01267, USA }}\vspace{3cm}

\end{center}
\subsection*{\centering Abstract}
{This paper defines a notion of parallel transport in a lattice
of quantum particles, such that the transformation associated
with each link of the lattice is determined by the quantum 
state of the two particles joined by that link.  We focus
particularly on a one-dimensional lattice---a ring---of 
entangled {\em rebits}, which are binary quantum objects confined
to a {\em real} state space.  We consider states of the ring that
maximize the correlation between nearest neighbors, and show that
some correlation must be sacrificed in order to have non-trivial
parallel transport around the ring.  An analogy is made with 
lattice gauge theory,
in which non-trivial parallel transport around closed loops is 
associated with a reduction in the {\em probability} of the field
configuration.  We discuss the possibility of extending
our result to qubits and 
to higher dimensional lattices.}

\vfill

PACS numbers: 03.67.-a, 11.15.Ha, 75.10.Jm

\newpage

\section{Introduction} 
In lattice gauge theory, the gauge field assigns to every pair of
neighboring lattice sites a transformation that tells how to
``transport'' a vector, representing an internal property
such as quark color, from one
site to the other.  That is, with every ordered pair 
$\langle j,k \rangle$ of neighboring
sites, or link, one associates a 
transformation $U(k,j)$ which is
an element of the gauge group.  Indeed, this assignment
of transformations to links {\em constitutes}
the configuration of the gauge field.  If we apply 
$U(k,j)$ to
a vector $v$ associated with site $j$, we can interpret the
image vector, $v' = U(k,j)v$, as the result of moving
$v$ from site $j$ to site $k$.\footnote{It helps to read the 
arguments of $U$ from right to left.  This notation makes sense
when combined with the usual notation for a sequence of
operations, in which the operator on the right acts first.}
This process is called parallel
transport, and the transformation $U(k,j)$ is sometimes called
a parallel transporter.

This paper is not about lattice gauge theory but about actual lattices
consisting of simple quantum particles, in which each
particle is correlated with its nearest neighbors.  As we will see below, for 
typical states of
the lattice, one can use the state itself to specify a notion of parallel
transport.   
Our conception is distinct 
from the more familiar notion of 
quantum parallel transport (expressible in terms
of gauge fields \cite{WZ,ACW}),
in which a quantum particle is physically moved either 
in actual space or in a parameter space \cite{ACW,Berry,AA}.
In our approach there is no physical motion or 
evolution; rather, the
transformation that we associate with a link expresses 
something about the relationship
between the particles joined by that link.
The present work is an initial exploration into a possible
analogy between the quantum state of a lattice of 
quantum particles and the configuration of the
gauge field in a lattice gauge theory.  
Roughly, the analogy we are looking for would be 
along the following lines (it will be spelled out more
precisely in later sections of the paper).

A lattice gauge theory assigns a probability distribution to the
set of possible field configurations, the probability density of 
a configuration ${\mathcal A}$ being proportional to 
$\exp[-S({\mathcal A})]$ where the real function $S({\mathcal A})$ 
is the action.  The action
depends on the effect of parallel transport around each of the
elementary plaquettes of the lattice; that is, it depends on the
transformations that the field configuration associates with these 
elementary closed paths.    
The more these transformations differ from the 
identity, the higher the action, and therefore the lower the 
probability of that particular field configuration.
Let us call this effect ``the probability
cost of twisting."  I am looking for something similar in 
a lattice of correlated quantum objects, but instead of a probability
cost, I am looking for a ``correlation cost.''  We will be focusing
our attention on states in which nearest neighbors are maximally
correlated.  The question is whether 
this maximal correlation 
must be reduced if there is to be non-trivial parallel transport
around closed loops (that is, transport whose net effect is not null),
and if so, whether the reduction in correlation becomes more severe
as the effect of such parallel transport differs further from 
the identity.  In other words, in an entangled lattice, is there
a correlation cost of twisting?

In fact this paper only begins to answer this question.  Although
we set up the problem for lattices of arbitrary dimension, the one concrete
example we work out in detail
is the case of a one-dimensional ring.  Moreover,
for most of the paper we will restrict our attention to the simplest 
possible quantum object, namely, a {\em rebit}, a fictitious object
whose state space is a two-dimensional vector space over the
{\em real} numbers \cite{rebit}.  Thus we will mostly be analyzing a 
closed ring of rebits.  At the end of the paper I discuss the generalization
to qubits and to higher-dimensional lattices.  

This research is related to a recent line of work 
on ``entanglement sharing," 
concerning the ways in which quantum 
entanglement can be shared among several objects.  A number of authors 
have found constraints on the sharing of entanglement that 
follow directly from the structure of quantum mechanics itself and not 
from any particular Hamiltonian.  For example, it has been shown that any 
entanglement that might exist between a pair
of qubits limits the extent to which either of them can be entangled
with a third qubit \cite{Bruss,CKW}.  There are similar limits for $n$ qubits 
all entangled with each other \cite{KBI,DVC,DW}.  
Another example is an entangled ring: in a translationally invariant
state of a ring of qubits, there is a certain maximum possible 
entanglement between nearest neighbors \cite{OW}.  
(The measure of entanglement
used in all of these studies is the entanglement of 
formation \cite{BDSW,BBPSSW}.) 
In this paper we are putting a 
different sort of
condition on the quantum state of a multipartite system---we are 
imposing a certain configuration of twists in the nearest-neighbor
correlations---and we are asking 
what
constraint this condition places on the strength of the 
correlations.  

I hope that the results of this research will ultimately be useful in
analyzing systems of entangled particles on a lattice, such as magnetic
systems.  If there are simple laws of quantum correlations that
transcend any particular Hamiltonian, then these laws might lead to
the identification of interesting generic properties of 
quantum many-body systems.  Arguments along these lines, particularly
focusing on entanglement, have appeared recently in the
literature \cite{ON1,Preskill,ZW,ON2}.  
But at least as much of the actual motivation for the present work
comes from pure curiosity: I wonder how close an analogy
one can draw between the degree of 
correlation between particles in a quantum lattice
and the probability density of a field configuration
in lattice gauge theory.  
Of course there are many
connections between lattice gauge theory and the theory of 
many-body systems---see, for example, Ref.~\cite{BGSS}
and references cited therein---but 
I am looking for an analogy along the particular lines
traced out above.   

The reader may have noticed that in describing the work to be 
presented here (as opposed to earlier work), I have been 
using the word ``correlation'' rather
than ``entanglement.''  Though they are related,
the two concepts are not the same.
In this paper I focus on correlation because it seems natural 
in this context and it is
easy to work with.
But it would also be interesting to 
explore the same questions using one of the standard
measures of entanglement.  I might add that the states we 
will primarily be concerned with are in fact highly entangled; 
hence the reference to an ``entangled ring" in the title.

The paper is organized as follows.  First we review briefly those
aspects of lattice gauge theory that have suggested 
our main question.  We then define a rebit
more precisely and develop our notion of parallel transport.
We analyze in some detail the case of a ring of rebits and 
determine whether non-trivial parallel transport does indeed entail
a ``correlation cost."  Finally we ask how the problem and the 
results are likely to change when extended to more complex
systems.  

\section{Lattice gauge theory and a simple analogy}

Ideally, lattice gauge theory is done on a four-dimensional
lattice representing spacetime, except that the fourth dimension 
represents
{\em imaginary} time, so that it acts in many respects like another
spatial dimension.  The results of a calculation can be interpreted
in terms of {\em real} time by means of analytic
continuation.  One consequence of the use of imaginary time is
this: in computing the expectation value of an observable, one 
does not sum up complex amplitudes associated with different histories;
rather, one takes a weighted average of the observable of interest using
{\em real} weights \cite{lattice}.  
As mentioned above, the weighting function is
proportional to $\exp(-S)$, where the action $S$ is a real function of the 
field configuration.  More precisely, in a pure gauge theory, 
in which there are
no matter fields but only the gauge field itself, 
the expectation value of an observable $B$
is
\begin{equation}
\langle B \rangle = \frac{1}{Z}\int B 
e^{-S({\mathcal A})} 
\prod_{\langle j,k \rangle}dU(k,j). \label{average}
\end{equation}
Here ${\mathcal A}$ is the configuration of the gauge
field, which assigns a parallel 
transporter $U(k,j)$ to each 
link $\langle j,k \rangle$.  
Each such transformation $U$ is an element of the gauge
group, {\em e.g.}, U(1) for electrodynamics or 
SU(3) for chromodynamics.  The integral in Eq.~(\ref{average})
is over all field configurations, that is, over all 
possible parallel transporters for each link, 
and $dU(k,j)$ indicates the 
invariant measure over the gauge group.  In the
integral, the parallel transporters for different
links are independent, except that $U(k,j)
=U(j,k)^{-1}$, so that only one of 
these two ordered pairs needs to be represented 
in the integral.  The normalizing constant $Z$ is
simply 
\begin{equation}
Z = \int 
e^{-S({\mathcal A})} \prod_{\langle j,k \rangle}dU(k,j).
\end{equation}

The action $S$ depends only on the results of parallel 
transport around plaquettes of the lattice,
{\em e.g.}, elementary squares in a cubic lattice. 
Let ${\mathcal P}$ be a plaquette, which we can think of as
a sequence of lattice sites; for a cubic lattice
${\mathcal P}$ would consist of four sites $j,k,l,m$.
Let $U({\mathcal P})$ be the net effect of parallel transport 
around plaquette ${\mathcal P}$; in the square example, 
$U({\mathcal P})$
would be $U(j,m)U(m,l)U(l,k)U(k,j)$ (the right-most operator acting first).
$S$ is defined so that it increases as the 
plaquette transformations $U({\mathcal P})$ get farther 
from the identity.  Various action functions with
this property have been used in the literature; the one originally
proposed by Wilson for an SU($N$) gauge field is \cite{Wilson}
\begin{equation}
S \propto -\sum_{\mathcal P}\,\hbox{Re(Tr}(U({\mathcal P}))),
\label{action}
\end{equation}
where the sum is over all plaquettes in the lattice.
Note that if ${\mathcal P}'$ consists of the same set of
points as ${\mathcal P}$, but with a different starting point or 
with the points taken in the opposite order, 
then $\hbox{Re(Tr}(U({\mathcal P}')))=
\hbox{Re(Tr}(U({\mathcal P})))$.
Thus we need include in the above sum only one ordered
set ${\mathcal P}$
representing each geometric plaquette.   

A gauge transformation associates with each lattice point
$j$ a group element $\Lambda(j)$, and under such a transformation
each parallel transporter $U(k,j)$ transforms
according to
\begin{equation}
U(k,j) \rightarrow \Lambda(k)U(k,j)
\Lambda(j)^{-1}.
\end{equation}
It is easy to see that, though a gauge transformation
changes the field configuration, it does not 
change any of the plaquette transformations 
$U({\mathcal P})$.  Therefore
it does not affect the action $S$ and so has no physical
consequences.  

This invariance under gauge transformations provides a 
simple analogy between the configuration of a gauge field
and the state of a lattice of quantum particles.  Consider,
for example, a lattice of qubits.  Rotating each of the 
individual qubits separately is analogous to a gauge
transformation.  The state of the lattice changes under
such rotations, but certain physical properties do not
change.  In particular, any reasonable measure of the 
degree of 
{\em entanglement} or {\em correlation} between two qubits does
not change.  So at least in this one modest respect, the degree of
correlation in a quantum lattice is similar to the action
or the probability density of a field configuration in 
a lattice gauge theory.  We want to see whether the similarity
goes any further than this.

\section{What is a rebit?}

\noindent As we have said, the quantum object we will mostly be
concerned with
in this paper is the {\em rebit}.  We now define this object
more precisely. 

A pure state $|\psi\rangle$ of a single rebit is simply a 
normalized vector in 
a two-dimensional real vector space.  A mixed state of a rebit is 
a mixture of pure states:
\begin{equation}
\rho = \sum_i p_i |\psi_i\rangle\langle\psi_i|,
\end{equation}
where $p_i > 0$ and $\sum_i p_i = 1$.  Equivalently, a mixed state 
can be represented as
a real, symmetric $2\times 2$ matrix with unit trace and no negative 
eigenvalues.  

Of course any rebit state is also a qubit state, and for our purposes
it will be helpful to think of rebits simply as restricted qubits.  
On the Bloch sphere, the restriction to real density matrices becomes a 
restriction to the $x$-$z$ plane.  But it will be more useful to 
change the representation by rotating the Bloch sphere.  Let us
rotate all states by 
$90^\circ$ in the left-handed sense around the positive $x$ axis,
so that our rebit states now lie in the $x$-$y$ plane.  
A general mixed state lying in this plane can be
written as 
\begin{equation}
\rho = \frac{1}{2}[I + a\sigma_x + b\sigma_y],
\end{equation}
where the $\sigma$'s are Pauli matrices and the real numbers 
$a$ and $b$ satisfy $a^2+b^2 \leq 1$.  Let us call this representation
of rebit states the ``horizontal representation," as opposed to the
original ``real-number representation."  Given a rebit density
matrix $\rho$ expressed in the horizontal representation, we can
always re-express it as a {\em real} density matrix $\rho_{real}$
simply by reversing the rotation around the $x$ axis:
\begin{equation}
\rho_{real} = U\rho U^\dag,
\end{equation}
where 
\begin{equation}
U = e^{-i(\pi/4)\sigma_x} = \frac{1}{\sqrt{2}}\mtx{cc}{1 & -i \\ -i & 1}.
\end{equation}
Here we have written the matrix in the basis
of eigenstates
of $\sigma_z$, 
$\{\mid\uparrow\rangle,\mid\downarrow\rangle\}$.
Note that $U\sigma_yU^\dag = \sigma_z$ and
$U\sigma_xU^\dag = \sigma_x$.

In the real-number representation, a pure state of $n$ rebits 
is a real vector in 
$2^n$ dimensions, and a mixed state of such a system is 
a real, symmetric $2^n \times 2^n$ density matrix.  
But again we will usually work in the horizontal representation, 
in which each rebit has been rotated by $90^\circ$
around the $x$ axis.  It will be helpful to have a simple way of recognizing
whether a given $n$-qubit state is a legitimate $n$-rebit state
expressed in the horizontal representation.  Conceptually, the test
is straightforward: apply the above transformation $U$ to each 
qubit---that 
is, rotate each qubit by $90^\circ$ (in the right-handed sense) around
the positive $x$ axis---and see whether the resulting state has only
real components.  Thus for a pure state $|\Psi\rangle$, we insist that 
$U^{\otimes n}|\Psi\rangle$ be real
in the standard up-down basis.  But
this is the same as saying that $U^{\otimes n}|\Psi\rangle =
(U^{-1})^{\otimes n}|\Psi^*\rangle$, where the asterisk indicates
complex conjugation in the standard basis.  Multiplying both
sides of this equation by $U^{\otimes n}$ and noting that $U^2 =
-i\sigma_x$, we arrive at the following criterion:
\begin{equation}
(-i\sigma_x)^{\otimes n}|\Psi\rangle = |\Psi^*\rangle.  \label{reality}
\end{equation}
In practice it will be simplest if we also allow ourselves to use
state vectors of the form $\exp(i\alpha)|\Psi\rangle$, where
$\alpha$ is real and $|\Psi\rangle$ satisfies Eq.~(\ref{reality}).
Though such state vectors do not become real when they 
are transformed by $U^{\otimes n}$, their density matrices do
become real. 
Allowing this possibility leads to the following weaker condition on 
an $n$-qubit
state $|\Psi\rangle$.
\begin{equation}
\sigma_x^{\otimes n}|\Psi\rangle = e^{i\beta}|\Psi^*\rangle ,
\label{propreal}
\end{equation}
$\beta$ being any real phase.  
Note that the matrix $\sigma_x$ simply 
interchanges $\mid\uparrow\rangle$ and
$\mid\downarrow\rangle$.  Thus we can recognize a pure $n$-qubit state
$|\Psi\rangle$ as a legitimate 
horizontal representation of an $n$-rebit state by
checking to see that the coefficient of each basis state, {\em e.g.},
$\mid\uparrow\uparrow\downarrow\uparrow\rangle$, is
the complex conjugate of the coefficient of the opposite state, in this case
$\mid\downarrow\downarrow\uparrow\downarrow\rangle$,
multiplied by a phase factor that is the same for all
basis states.  
Let us call Eq.~(\ref{propreal}) the ``rebit condition" for
pure states.  The corresponding
test for
mixed states can be obtained by a similar argument; one finds that a 
density matrix $\rho$ of $n$ qubits is the horizontal representation
of a legitimate $n$-rebit state if and only if
\begin{equation}
\sigma_x^{\otimes n}\rho \sigma_x^{\otimes n} = \rho^*,
\label{mixedreal}
\end{equation}
the complex conjugation again being in the standard basis.

We conclude this section with a word about {\em rotations} of a rebit.
Viewing the states of a rebit as qubit states confined to
the equatorial plane of the Bloch sphere, we could take as the allowed
rotations all the unitary transformations that represent
rotations around the $z$ axis, that is, all transformations
of the form
\begin{equation}
R = \mtx{cc}{e^{i\alpha} & 0 \\ 0 & e^{i\beta}}.
\end{equation}
However, our definition of
parallel transport will not be able to distinguish 
unitary transformations that are different only by
an overall phase factor; so we will call such transformations
identical.  For definiteness we pick a standard representative
from each of the resulting equivalence classes: a rotation
by an angle $\xi$ around the $z$ axis, with $0 \leq \xi < 2\pi$, 
will be represented by the matrix
\begin{equation}
R = \mtx{cc}{1 & 0 \\ 0 & e^{i\xi}}. \label{rot}
\end{equation}

\section{Parallel transport in a lattice of rebits}\label{parallel}

\noindent We now consider a lattice of rebits, on which we want to define
a notion of parallel transport.  For now the structure
of the lattice does not matter, as long as any two rebits are identified 
as either being neighbors or not.  Thus the lattice is
simply a graph.  The mathematical object that is to be parallel
transported is a pure state of a single rebit, which we can picture
as a direction in the horizontal plane.  The parallel transporter 
associated with a link in the lattice will be a rebit rotation---represented 
in the form (\ref{rot})---which we can identify with an 
element of U(1).  We want the
assignment of parallel transporters to links to be determined
by the quantum state of the lattice itself.  

We begin with the following scenario. 
Consider a pair of neighboring
rebits labeled $j$ and $k$.  They are in some joint state $\rho^{(jk)}$, 
a $4\times 4$ density matrix obtained from the 
state of the whole lattice by tracing over all the other particles.
Imagine performing an arbitrary
orthogonal measurement on particle $j$.  In the standard von Neumann
model this measurement brings particle $j$ into a certain pure
state---one of the eigenstates of the measurement---and it also brings
particle $k$ into some state, typically a mixed state, determined as
follows.  Let $|\psi^{(j)}\rangle$ be the
eigenstate into which particle $j$ is brought by the measurement.  Then
the post-measurement state of particle $k$ is the $2 \times 2$ density
matrix 
\begin{equation}
\omega^{(k)} = \frac{1}{P}\langle\psi^{(j)}|\rho^{(jk)}|\psi^{(j)}\rangle, 
\label{map}
\end{equation}
where $P = 
\langle\psi^{(j)}|\hbox{Tr}_k\rho^{(jk)}|\psi^{(j)}\rangle$ is the 
probability with
which that particular outcome occurs.  In Eq.~(\ref{map}) 
the matrix operations are done only in the space of
particle $j$.  To be more explicit, if $r=0,1$
and $s=0,1$ are indices associated with particles $j$ and $k$
respectively, we can write the components of $\omega^{(k)}$ in terms of
the components of $|\psi^{(j)}\rangle$ and $\rho^{(jk)}$ as 
\begin{equation}
\omega^{(k)}_{ss'} =\frac{1}{P} \sum_{r,r'}{{\psi}_r^{(j)}}^*
\rho^{(jk)}_{rs,r's'}\psi^{(j)}_{r'}. 
\end{equation}

We want to use this measurement scenario to associate
with the two-particle state
$\rho^{(jk)}$ a simple rotation $U(k,j)$.
First, let ${\mathcal M}$ (for ``measurement'') be the mapping defined by 
Eq.~(\ref{map}), which takes
each pure state of particle $j$
for which $P\neq 0$ into a pure or mixed state of 
particle $k$; that is, ${\mathcal M}(\psi^{(j)}) = \omega^{(k)}$. 
If $P=0$, let us say for definiteness that ${\mathcal M}(\psi^{(j)}) = 
\hbox{Tr}_j \rho^{(jk)}$, though it will not actually matter in what
follows.
For an arbitrary rebit rotation $R$, we define a function 
$F(R)$ by
\begin{equation}
F(R) = \frac{\int \langle \psi | R^{\dag} 
{\mathcal M}(\psi) R | \psi\rangle Pd\psi}
{\int P d\psi}.
\label{F}
\end{equation}
Here $P$ is the probability given above, and $d\psi$ indicates the
uniform measure over the circle of pure rebit states, normalized
so that $\int d\psi = 1$.  Thus $F(R)$
is an average fidelity of ${\mathcal M}(\psi)$, 
not with respect to $|\psi\rangle$
itself but with respect to a rotated version of $|\psi\rangle$.  
As we will see in the next paragraph, 
depending on the density matrix $\rho^{(jk)}$, one of the following
two conditions will hold: (i) $F(R)$ is independent of $R$, or 
(ii) there is a unique rotation
$R = U$ that maximizes $F(R)$.  In case (i), we say that there is no
correlation between particles $j$ and $k$.  In case (ii), we take the
special rotation $U$ that maximizes $F(R)$
to be the parallel transporter associated with
the link $\langle j,k \rangle$.  In a certain sense, 
$U$ is the rotation that most closely approximates
the action of ${\mathcal M}$. 

Combining Eqs.~(\ref{map}) and (\ref{F}) and the definition of 
$P$, we have
\begin{equation}
F(R) = \frac{\int (\langle\psi |\otimes\langle\psi |R^\dag )
\rho^{(jk)} (|\psi\rangle \otimes R|\psi\rangle ) d\psi}
{\int \langle \psi | \hbox{Tr}_k \rho^{(jk)} |\psi\rangle d\psi}.
\label{Fagain}
\end{equation}
It is not hard to show that the denominator is always 1/2.
Also, 
in the horizontal representation, the integral in the numerator
involves rotating $|\psi\rangle$ around the $z$ axis, so that we can 
rewrite Eq.~(\ref{Fagain}) as
\begin{equation}
F(R) = 2 
(\langle\psi_0 |\otimes\langle\psi_0 |R^\dag )
\Bigg[\frac{1}{2\pi}\int_0^{2\pi}
e^{i\gamma S_z} \rho^{(jk)} e^{-i\gamma S_z}d\gamma \Bigg]
(|\psi_0\rangle \otimes R|\psi_0\rangle )  ,
\label{Fnext}
\end{equation}
where $S_z = (\sigma_z^{(j)}+\sigma_z^{(k)})/2$ and
$|\psi_0\rangle$ is some fixed reference state which for definiteness
we take to be $|\psi_0\rangle = (\mid\uparrow\rangle
+\mid\downarrow\rangle )/\sqrt{2}$.
Now, a general two-particle density matrix satisfying the
rebit condition (\ref{mixedreal}) is of the form
\begin{equation}
\rho^{(jk)} = \mtx{cccc}{a&x_1&x_2&x_3\\x_1^*&b&c&x_2
\\x_2^*&c^*&b&x_1\\x_3^*&x_2^*&x_1^*&a}, \label{rho}
\end{equation}
the representation being in the
standard basis $\{\mid\uparrow\uparrow\rangle,
\mid\uparrow\downarrow\rangle, 
\mid\downarrow\uparrow\rangle,
\mid\downarrow\downarrow\rangle\}$.
The average over $\gamma$ in Eq.~(\ref{Fnext}) has the effect
of replacing
the $x_i$'s in $\rho^{(jk)}$ 
with zero and leaving the matrix elements
$a$, $b$, and $c$ unchanged.  
To evaluate $F(R)$, we write $R$ explicitly as a rotation around the
$z$ axis by some angle $\xi$:
\begin{equation}
R = \mtx{cc}{1 & 0 \\ 0 & e^{i\xi}} .
\label{Rxi}
\end{equation}
Inserting this matrix into Eq.~(\ref{Fnext}) we find that
\begin{equation}
F(R) = \frac{1}{2}[ 1 + 2|c| \cos(\xi - \phi) ],
\label{FR}
\end{equation}
where $\phi$ is the phase of the matrix element
$c$; that is, $c = |c|e^{i\phi}$.
If $c$ is zero, then we find ourselves in 
case (i) mentioned above: $F(R)$
is independent of $R$.  Otherwise $F$ is maximized when
the angle of rotation $\xi$ is equal to the phase $\phi$.  
According to our prescription,
then, the parallel transporter $U(k,j)$ is the rotation given
in Eq.~(\ref{Rxi}) with $\xi = \phi$.   

We will take as our measure of the degree of correlation 
between particles $j$ and $k$ the quantity $2|c|$, which 
ranges from 0 to 1.  Eq.~(\ref{FR}) makes it clear that
$2|c|$ measures the degree of angular correlation between
the two particles.  We can also interpret this quantity
in terms of the more standard correlation matrix
$\tau_{\mu\nu} = \hbox{Tr}\big[\rho^{(jk)}(\sigma_\mu \otimes
\sigma_\nu)\big]$, in which both $\mu$ and $\nu$ take as values the
axis labels $x$ and $y$.  One finds that $(2|c|)^2 = 
[(\tau_{xx}+\tau_{yy})^2 + (\tau_{xy}-\tau_{yx})^2]/4=
[2\det\tau + \hbox{Tr}(\tau^T\tau)]/4$.

Note that the single complex number $c$ determines both
the parallel transporter (through its phase) and the degree
of correlation (through its magnitude).  Typically we will be
trying to maximize the magnitude of $c$ for a given value of its phase.
The fact that it is possible for a link $\langle j,k \rangle$ to
have an undefined parallel 
transporter $U(k,j)$ (case (i) above, where
$c=0$)
will not cause any difficulties for the problem we will be
studying.  We will be considering the set of 
all states that are consistent
with a given specification of the parallel transporters, {\em i.e.},
the phases of the $c$'s.
If a link has $c=0$, then we simply say that that link is
consistent with any specified phase.

Our notion of parallel transport has a particularly simple
interpretation if $\rho^{(jk)}$ has the form (\ref{rho})
with all the $x_i$'s equal to zero.  This will happen, for 
example, if the state of the lattice is invariant under 
identical rotations of all the rebits.  If $\rho^{(jk)}$
has this form, then Eq.~(\ref{map}) yields
\begin{equation}
\omega^{(k)} = p\,U(k,j)|\psi^{(j)}\rangle\langle\psi^{(j)}| U(k,j)^\dag
+ (1-p) (I/2),
\end{equation}
where $I$ is the $2\times 2$ identity and $p = 2|c|$.
Thus the post-measurement state of particle $k$ is
simply a rotated and partially depolarized version of 
$|\psi^{(j)}\rangle$, and the weight $p$ of the 
pure rotated state is our measure of correlation.   

Let us now imagine 
transporting, mathematically, a rebit state around a closed loop 
in accordance with the above prescription.  As we will see,
the final state in such a process need not be the same as the initial 
state but could
be rotated by some angle $\theta$.
One might ask:  How does one interpret physically 
this process of transport,
and what is the meaning of the rotation angle $\theta$?  I regard
our concept of parallel transport primarily as a mathematical
notion; nothing is being physically transported.  However, in the
special case considered above, in which the state of the lattice
is rotationally invariant, one can 
extract from our definition a simple
physical interpretation of the net rotation angle.  Consider a 
closed loop of $n$ lattice sites $j_0, j_1, \ldots, j_{n-1}$.
At each of the sites $j_0, \ldots, j_{n-2}$, that is, at all
but the last site, perform 
an orthogonal measurement,
with outcomes labeled `0' and `1', choosing the measurement
at $j_m$ so that it maximizes the probability of getting the
same outcome (0 or 1) as at site $j_{m-1}$.  That is,
we are trying to minimize the expected number of 
flips from 0 to 1 or from 1 to 0 as we go around
the loop.  Now, at the 
last site, $j_{n-1}$, one is faced with a dilemma: there 
will be a measurement that maximizes the probability
of agreement between $j_{n-2}$ and $j_{n-1}$, and there
will be a (possibly different) measurement that maximizes
the probability of agreement between $j_{n-1}$ and $j_0$.
The angle between these two measurements, that is, between
their `0' eigenstates, is the angle $\theta$ associated with parallel
transport around the loop.  We can think of this angle as
measuring the net ``twist'' in the nearest-neighbor correlations.

When the state of the lattice is not rotationally invariant,
the angle between the two competing
optimal measurements at the last
site (the two being reckoned optimal from different directions) 
may depend on the choice of the initial measurement on
the first particle.  So the interpretation in this case 
is not as simple.
Still, it is reasonable to think of the net rotation angle as
a measure of the net twist in the correlations.  
In the following section where 
we analyze the case of a rebit ring, we will find that the optimal
states, which are the states most relevant to our problem, 
are in fact rotationally invariant,
so that the above interpretation applies.    

\section{Analysis of a rebit ring} \label{ring}

So far we have not made any assumptions about the structure of the
lattice.  In order to obtain a concrete result, we now specialize 
to the simplest possible lattice for which our general question
can be addressed, namely, a closed one-dimensional ring.  
Let the ring consist of $n$ rebits with $n\geq 2$, 
labeled by $j = 0, 1, \ldots$, 
\hbox{$n-1$}; the labeling is mod $n$, so that $j=n$ is the 
same as $j=0$.
Let $c_j$ be the matrix
element $c$ of Eq.~(\ref{rho}) when the two particles in question are
particles $j$ and $j+1$.  
The quantity $K = \frac{2}{n}\sum_j |c_j|$ will be the
measure of average nearest-neighbor correlation
that we will be trying to maximize; note that
$0 \leq K \leq 1$.   The product 
$\prod_j (c_j/|c_j|)$, which we will 
call $e^{i\theta}$, is the net phase factor 
associated
with transport around the whole ring, and $\theta$ (defined 
only mod $2\pi$) is the net
rotation angle.  Our question is this: what is the
maximum possible value of $K$ for a fixed value of $\theta$?  Let
us call this maximum value $K_{\max}(\theta)$.  If 
$K_{\max}(\theta)$ decreases as $e^{i\theta}$ gets farther from unity,
then we can say that there is a correlation cost associated with
non-trivial parallel transport.

As we have stated the problem so far, the phases of the different
$c_j$'s, that is, the phases that define the individual parallel
transporters, need not be the same for all links in the ring.  However,
for our purpose there is no loss of generality in assuming that these
phases are all equal.  This is because if they were not equal, we could
always apply local rotations to the individual rebits (analogous to 
a gauge transformation) so as to make them
equal.  Local rotations can change neither the magnitude of any
$c_j$ nor the overall phase factor $e^{i\theta}$.  So the restriction
to equal phases does not eliminate any states that might change the
answer to our question.  That is, states with maximal $K$ for any given
$\theta$ are still represented in the restricted set.

With this restriction, we can simplify our problem by expressing the
average correlation $K$ and the overall phase $\theta$ in terms of
creation and annihilation operators.  Let $a_j$ be an operator on
particle $j$ 
defined by $a_j\hspace{-1.5mm}\mid\uparrow\rangle = 
\mid\downarrow\rangle$ and
$a_j\hspace{-1.5mm}\mid\downarrow\rangle = 0$.  
Then for the link between the $j$th and
the $(j+1)$st rebits, we can write $c_j$ as
\begin{equation}
c_j = \hbox{Tr}(\rho a^\dag_{j+1} a_{j}),
\end{equation}
$\rho$ being the density matrix of the ring.
Now if we define the operator $\Gamma$ to be  
\begin{equation}
\Gamma = \frac{1}{n}\sum_{j=0}^{n-1}a^\dag_{j+1} a_{j}
\end{equation}
and write $\langle\Gamma\rangle = \hbox{Tr}\,\rho\Gamma$,
it follows (assuming that each $c_j$ has the same phase) that
\begin{equation}
K =2|\langle\Gamma\rangle|,  \label{CG}
\end{equation}
and that the overall phase $\theta$ (mod $2\pi$) is simply 
\begin{equation}
\theta = n\, \hbox{arg}\big( 
\langle\Gamma\rangle\big). \label{theta}
\end{equation}  
$\Gamma$ is not a
Hermitian operator---so its eigenvalues may be complex and indeed must be
complex if $\theta$ is to take any non-trivial value---but $\Gamma$ does
commute with its adjoint, which implies that eigenvectors
corresponding to distinct eigenvalues are orthogonal.  
Note also that $\Gamma$ commutes with the total $z$ component
of spin; so the eigenstates of $\Gamma$ can be taken to be
states with a definite number $u$ of up spins.  Our immediate
goal is to find the eigenvalues of $\Gamma$, from which we 
will be able to determine the set of possible values of
$\langle \Gamma \rangle$, which in turn will give us 
$K_{\max}(\theta)$. 

The $a$ operators are not quite fermionic creation and annihilation
operators, because the operators associated with different sites commute
with each other rather than anticommute.  However, 
we can use a standard
trick \cite{LSM} to define genuinely fermionic operators $b_j$:
\begin{equation}
b_j = \exp\bigg[i\pi\sum_{k=0}^{j-1}a_k^\dag a_k\bigg] a_j.
\end{equation}
The operators $b_j$ satisfy the usual fermionic anticommutation relations:
\begin{equation}
\{b_j,b_k\} = \{b^\dag_j,b^\dag_k\} = 0; \,\,\,\{b_j,b^\dag_k\} = \delta_{jk}.
\end{equation}
We now express $\Gamma$ in terms of the $b$'s.  The expression
depends on whether $u$, the number of up spins in the
ring, is even or odd; that is, the expression is different in 
different subspaces.
For odd values of $u$, $\Gamma$ looks the same in terms 
of the $b$'s as it does
in terms of the $a$'s:
\begin{equation}
\Gamma = \frac{1}{n}
\sum_{j=0}^{n-1}b^\dag_{j+1} b_{j}.
\end{equation}
 For even values of $u$, there is
a sign change in the last term:
\begin{equation}
\Gamma = \frac{1}{n}\Bigg[
\sum_{j=0}^{n-2}b^\dag_{j+1} b_{j} - b_{0}^\dag b_{n-1}\Bigg].
\end{equation}
In either case, we can
diagonalize $\Gamma$ and find its exact single-fermion eigenvalues. 
From these we can obtain the eigenvalues of 
$\Gamma$ for an arbitrary value of $u$ 
by summing $u$ of the single-fermion eigenvalues. 

For odd $u$, the single-fermion eigenstates of 
$\Gamma$ are $|\omega_m\rangle = d_m^\dag |0\rangle$,
$m= 0, \dots, n-1$,
where $|0\rangle$ is the vacuum state, that is, the 
state with all spins down, and the creation operator $d_m^\dag$
is given by
\begin{equation}
d_m^\dag = \frac{1}{\sqrt{n}}\sum_{j=0}^{n-1} e^{-2m\pi i j/n} b_j^\dag.
\end{equation}
The corresponding eigenvalues of $\Gamma$ are 
\begin{equation}
g_m = (1/n)e^{2m\pi i/n}.
\end{equation}
For even $u$, the creation operators for the single-fermion
eigenstates are
\begin{equation}
d_m^\dag = \frac{1}{\sqrt{n}}\sum_{j=0}^{n-1} 
e^{-(2m+1)\pi i j/n} b_j^\dag,
\hspace{0.7cm} m = 0, \ldots, n-1,
\end{equation}
and the corresponding eigenvalues are 
\begin{equation}
g_m = (1/n)e^{(2m+1)\pi i/n}.
\end{equation}
Regardless of the value of $u$, the single-fermion eigenvalues
of $\Gamma$ are complex numbers of length $1/n$, 
with phases uniformly spaced
around the complex plane.  

The eigenvalues
of $\Gamma$ corresponding 
to a system of $u$ fermions (that is, $u$ up spins) are
all the possible sums of $u$ of the $g_m$'s.
That is, we can write each such eigenvalue as
\begin{equation}
G(M) = \sum_{m\in M} g_m,  \label{gG}
\end{equation}
where $M$ is a set of exactly $u$ integers chosen from 
the set $\{0, \ldots, n-1\}$.
(We cannot use the same value of $m$ twice in this sum because 
no two identical fermions can be in the same state, and all the
single-fermion eigenvalues are non-degenerate.)

As we will see, the most important eigenvalues $G(M)$ for
our purpose will be the ones with the greatest magnitude.
For even values of $n$,
these are the ones for which $u=n/2$ and the set $M$ consists of 
a string of consecutive integers (mod $n$), so that the
corresponding values $g_m$ constitute a ``fan'' of 
complex numbers spread out over half of the complex
plane.   Performing the sum in Eq.~(\ref{gG}), we find 
that these extreme eigenvalues are
\begin{equation}
G_r = 
\frac{e^{2r\pi i/n}}{n\sin(\pi/n)},\hspace{0.7cm}r = 0, 
\ldots, n-1, \hspace{0.6cm} \hbox{(even $n$)}
\label{ngon}
\end{equation}
an equation that holds for all even values of $n$ (even
though $u=n/2$ might be even or odd).  
For odd values of $n$, the eigenvalues with largest magnitude are obtained
by setting $u$ equal to either $(n+1)/2$ or $(n-1)/2$, and again
letting $M$ consist of a string of consecutive integers mod $n$.
For either of these choices of $u$, the eigenvalues
thereby obtained are
\begin{equation}
G_r = \frac{e^{2r\pi i/n}}{n\sin(\pi/n)}\cos(\pi/(2n)),
\hspace{0.7cm}r = 0, \ldots, n-1.\hspace{0.6cm}\hbox{(odd $n$)}
\label{odd}
\end{equation}

For either even or odd $n$, each eigenstate corresponding 
to one of the eigenvalues $G_r$ can be
written as 
\begin{equation}
|\Omega_{r,u}\rangle = d^\dag_{m_0}d^\dag_{m_0+1}\cdots
d^\dag_{m_0+u-1}|0\rangle,
\end{equation}
where $m_0 = r-\lfloor u/2 \rfloor$ mod $n$.  (Here and below,
addition in the subscript of $d$ is always mod $n$.)  
This value of $m_0$
places $\exp(2r\pi i /n)$ in the center of the fan of complex
eigenvalues $g_{m_0}, \ldots, g_{m_0+u-1}$.  For later convenience,
we define the following density matrices based on these eigenstates.
For even $n$,
\begin{equation}
\rho_r = |\Omega_{r,n/2}\rangle\langle\Omega_{r,n/2}|,
\end{equation}
and for odd $n$,
\begin{equation}
\rho_r = \frac{1}{2}\bigg[|\Omega_{r,(n+1)/2}\rangle
\langle\Omega_{r,(n+1)/2}| +|\Omega_{r,(n-1)/2}\rangle
\langle\Omega_{r,(n-1)/2}|\bigg].
\end{equation}
As we will see shortly, both of these density matrices satisfy the
rebit condition, even though $|\Omega_{r,(n+1)/2}\rangle$
and $|\Omega_{r,(n-1)/2}\rangle$ do not.  Note also that
for both even and odd $n$,
$\hbox{Tr}\,\rho_r \Gamma = G_r$.

One can check that $\rho_r$ 
is translationally invariant and thus is in accord with the assumption
we made earlier, that the matrix element $c_j$ has the same
phase for each link of the ring.  But we also want to check 
that $\rho_r$ is a legitimate 
rebit state, {\em i.e.}, that it satisfies Eq.~(\ref{mixedreal}).
Let us do this first for even values of $n$, in which case we
are dealing with a pure state $|\Omega_{r,n/2}\rangle$.

We begin by noting that 
\begin{equation}
\sigma_x^{\otimes n}b_j^{\dag} = (-1)^j b_j \sigma_x^{\otimes n},
\end{equation}
which can be seen directly from the definition of $b_j$.
It follows that 
\begin{equation}
\sigma_x^{\otimes n}d_{m_1}^\dag\cdots d_{m_u}^\dag |0\rangle
= \Big[ d_{m_1 + n/2} \cdots d_{m_u + n/2}\sigma_x^{\otimes n}
|0\rangle\Big]^*,   \label{ds}
\end{equation}
where $m_1, \ldots, m_u$ are any distinct values chosen
from the set $\{0,\ldots, n-1\}$.  (The addition of $n/2$ in the
subscripts on the right-hand side comes from $(-1)^j$ in the
preceding equation.)
In general, a state of the form 
$d_{m_1}^\dag\cdots d_{m_u}^\dag |0\rangle$ will not
satisfy the rebit condition, even if $u=n/2$.  However, for
the special case in which $m_1, \ldots, m_u$ are $n/2$
{\em consecutive} integers, as they are in the definition
of $|\Omega_{r,n/2}\rangle$, the subscripts on the right-hand side
of Eq.~(\ref{ds}) are precisely those elements of
$\{0,\ldots, n-1\}$ that are not included in $\{m_1,\ldots,m_u\}$.
Therefore, when those annihilation operators are applied to
the all-up-spin state $\sigma_x^{\otimes n}|0\rangle$,
which within a phase factor is the same as
$d_0^\dag d_1^\dag \cdots d_{n-1}^\dag |0\rangle$,
the resulting state, again up to an overall phase factor, is  
$d^\dag_{m_1}\cdots d^\dag_{m_u}|0\rangle$.
We have thus shown that 
\begin{equation}
\sigma_x^{\otimes n}
|\Omega_{r,n/2}\rangle = e^{i\beta}|\Omega_{r,n/2}^*\rangle
\end{equation}
for some phase $\beta$, so that 
$\rho_r$ satisfies the rebit condition (\ref{propreal}) for even values of $n$.  

Turning now to the case of odd $n$, one can use an argument like the one
in the preceding paragraph to show that 
\begin{equation}
\sigma_x^{\otimes n}|\Omega_{r,(n-1)/2}\rangle
\langle\Omega_{r,(n-1)/2}|\sigma_x^{\otimes n}
=|\Omega_{r,(n+1)/2}^*\rangle
\langle\Omega_{r,(n+1)/2}^*|,
\end{equation}
and vice versa, so that
\begin{equation}
\sigma_x^{\otimes n}\rho_r \sigma_x^{\otimes n}
=\rho_r^*.
\end{equation}
So $\rho_r$ satisfies the rebit condition for odd values
of $n$ as well.

We are now in position to find the 
set---call it ${\mathcal G}$---of possible values 
of $\langle \Gamma \rangle$, from which we will be able to
determine $K_{\max}(\theta)$.  
The complex numbers $G_r$ given by Eq.~(\ref{ngon}) or Eq.~(\ref{odd}), 
being 
values of $\langle\Gamma\rangle$ corresponding to 
the legitimate 
rebit states $\rho_r$, are elements of
${\mathcal G}$.  By taking mixtures of
these states, we can obtain other possible values of 
$\langle\Gamma\rangle$.  Let
\begin{equation}
\rho = \sum_r q_r \rho_r,
\end{equation}
where 
the $q$'s are non-negative numbers summing to 1.
For this state we have
\begin{equation}
\langle\Gamma\rangle = \,\hbox{Tr}\,\rho\Gamma = \sum_r q_r G_r.
\label{mixture}
\end{equation}
The complex numbers $G_r$ are the vertices of a
regular $n$-gon in the complex plane, and Eq.~(\ref{mixture})
shows that this $n$-gon and its interior are contained in
${\mathcal G}$.  

In fact it is easy to see that ${\mathcal G}$
contains no other points.  Any complex number 
$\langle\Gamma\rangle$ in ${\mathcal G}$
must be a weighted average of eigenvalues of $\Gamma$:
\begin{equation}
\langle\Gamma\rangle = \sum_M q_M G(M).
\end{equation}
But one can show that each eigenvalue $G(M)$, regardless
of the value of $u$, lies on or inside the $n$-gon defined by the
special eigenvalues discussed above.  Therefore it is impossible
for the average to get outside this region.

For the special case $n=2$, the interior of 
the ``$n$-gon'' is simply a
segment of the real axis, running from $-1/2$ to $+1/2$.
Thus the only possible phases of $\langle\Gamma\rangle$ are 
zero and $\pi$, so that according to Eq.~(\ref{theta})
the only possible value of $\exp(i\theta)$ is 1.  (This is
because there is no real loop to traverse; to return
to the starting place, one has to retrace one's steps.) 
For all other values of $n$, all values of $\theta$
from 0 to $2\pi$ are possible.  To see this, it is enough
to consider a single side of the $n$-gon; let us take the side consisting
of the line segment joining the point 
$G_0$, on the positive real axis, with the point
$G_1 = e^{2\pi i/n}G_0$.
As we travel along this segment, the phase of
$\langle\Gamma\rangle$ varies from $0$ to
$2\pi/n$, so that $\theta$ varies from $0$ to
$2\pi$.  The range of values of $|\langle\Gamma\rangle |$
as a function of $\theta$ is the same for each of 
the other sides of the $n$-gon.

It thus becomes a simple geometric problem 
to find $K_{\max}(\theta)$.  For $n>2$, consider the
line segment just described, 
connecting $G_0$ to $G_1$, and note that for any $\theta$ in the range
$0\leq\theta\leq 2\pi$, 
$K_{\max}(\theta)$ is twice the magnitude of the unique
point along this segment whose phase is
$\theta/n$.  Doing the geometry, and using the values of
$G_r$ given in Eqs.~(\ref{ngon}) and (\ref{odd}), one finds
that for even $n$,
\begin{equation}
K_{\max}(\theta) = \frac{2\cos(\pi/n)}{n\sin(\pi/n)\cos[(\pi-\theta)/n]},
\label{answer}
\end{equation}
and for odd $n$,
\begin{equation}
K_{\max}(\theta) = \frac{2\cos(\pi/n)\cos(\pi/(2n))}
{n\sin(\pi/n)\cos[(\pi-\theta)/n]}.
\label{oddanswer}
\end{equation}
It is clear both from the geometric picture and from Eqs.~(\ref{answer})
and (\ref{oddanswer})
that $K_{\max}$
is largest at $\theta=0$ and $\theta=2\pi$ and 
smallest at $\theta =\pi$.  Indeed, the value of $K_{\max}$
becomes smaller the more $e^{i\theta}$ differs from unity.
In this sense there is a correlation cost of non-trivial 
parallel transport around the ring.  

It is not hard to interpret the states $\rho_r$ 
physically.  The state $\rho_0$, which entails no twisting
as one goes around the ring, is the ground
state, or in the case of odd $n$ an equal mixture of the
two degenerate ground states, of the ferromagnetic 
$XY$ model \cite{LSM} on a
one-dimensional ring, whose 
Hamiltonian is $H = -\sum_j (a_j^\dag a_{j+1} + a_{j+1}^\dag a_j)
=-n\big(\Gamma^\dag + \Gamma\big)$.  
The state $\rho_r$ can be obtained from 
$\rho_0$ by rotating each rebit, the rotation angle 
at site $j$ being $2\pi rj/n$.
These rotations do not change the strength of the 
nearest-neighbor correlations, but for each site $j$ they change the
phase of the matrix element $c_j$ 
from zero to $2\pi r/n$.
Still, this does not change the {\em overall} phase 
factor $e^{i\theta}$ associated with 
the whole ring.  When one creates a mixture of 
two of these differently rotated states, {\em e.g.},
$\rho_0$ and $\rho_1$, the resulting matrix element 
$c_j$ is an average of two complex numbers with different
phases.  It is this averaging process that allows
the possibility of a non-trivial net phase change around 
the ring.

In the case of even $n$, where $\rho_r$ represents the
pure state $|\Omega_{r,n/2}\rangle$, one could achieve
the same averaging effect by creating coherent 
superpositions of eigenstates of $\Gamma$ rather 
than incoherent mixtures.  I have chosen to use mixtures
because superpositions of these eigenstates are not
necessarily translationally invariant.  However, just to 
demonstrate that it is possible for a translationally
invariant pure state to have a non-zero
rotation angle $\theta$ associated with parallel transport
around the ring, I offer the following example for $n=6$:
\begin{equation}
|\Psi\rangle = \frac{i}{\sqrt{12}}
\Big[e^{i\phi/2}(\mid\uparrow\uparrow\downarrow
\uparrow\downarrow\downarrow\rangle + \cdots )
+e^{-i\phi/2}(\mid\uparrow\uparrow\downarrow
\downarrow\uparrow\downarrow\rangle + \cdots )
\Big].  \label{six}
\end{equation}
Here each ellipsis stands for all possible translations
of the given state.  (Though we are not paying particular 
attention to the overall phase factors of pure states, I 
have chosen the overall phase in Eq.~(\ref{six}) 
to satisfy Eq.~(\ref{reality}).)  
One finds that the matrix element $c_j$ for each
link in this ring is $(1/6)\exp(i\phi)$, so that the overall
rotation angle is $\theta = 6\phi$ and the average 
correlation is $K = 1/3$.  Thus any value of $\theta$
can be realized with a translationally invariant pure state.  
But the value $K=1/3$ is not optimal.  
To obtain the optimal value $K_{\max}(\theta)$
in a translationally invariant state, one must typically
use mixed states rather than pure states.

We finish this section by giving asymptotic expressions
for $K_{\max}(\theta)$ as the number of rebits in the
ring gets very large.  For even $n$, Eq.~(\ref{answer})
to order $1/n^2$ becomes (for $0 \leq \theta \leq 2\pi$)
\begin{equation}
K_{\max}(\theta) = \frac{2}{\pi}\bigg[1+\frac{1}{n^2}
\bigg(\frac{\pi^2}{6}-\pi\theta+\frac{\theta^2}{2}\bigg)\bigg],
\label{evenasymp}
\end{equation}
while for odd $n$ we have
\begin{equation}
K_{\max}(\theta) = \frac{2}{\pi}\bigg[1+\frac{1}{n^2}
\bigg(\frac{\pi^2}{24}-\pi\theta+\frac{\theta^2}{2}\bigg)\bigg].
\label{oddasymp}
\end{equation}
Thus the correlation cost of non-trivial parallel transport
becomes smaller as the size of the ring increases.

\section{Other lattices}

Let us now think about how the above problem might be
generalized to a finite or infinite lattice of higher dimension (still
using rebits as our basic objects).  We can state the 
problem as follows.  As in Section \ref{parallel},
let $\rho^{(jk)}$ be the density matrix of the pair of rebits
at the neighboring sites $j$ and $k$, and let $c_{j,k}$ be
the coefficient of $\mid\uparrow\downarrow\rangle
\langle\downarrow\uparrow\mid$ in this density matrix.
(This $c_{j,k}$ is analogous to the $c_j$ of the preceding
section.)  Now suppose that the phases of all the $c$'s,
for all the links $\langle j,k \rangle$, are specified.  We 
will call this complete specification ${\mathcal A}$, since
it is analogous to a field configuration
in the U(1) gauge theory.  Given this specification, we
have two questions: (i) Is it possible to find a lattice 
state for which the numbers $c_{j,k}$ are all non-zero and
have the chosen phases?  (One can always find a state in which all the
$c_{j,k}$'s are zero, making the state consistent with any phases, 
but such a state is not very interesting.)  (ii) What is the 
maximum possible value of 
\begin{equation}
K = \frac{2}{L}\sum_{\langle j,k \rangle} |c_{j,k}|,
\end{equation}
consistent with the specification ${\mathcal A}$?  
Here $L$ is the number of links in the lattice.  In the case
of an infinite lattice, $K$ can be defined as a limit over
a sequence of finite lattices. 
Let us call the
maximum value $K_{\max}({\mathcal A})$.

For definiteness let us consider a specific lattice, namely,
an infinite square lattice in two dimensions.  Let us consider
first the configuration ${\mathcal A}_0$ in which the phases
of all the $c_{j,k}$'s are zero.  
In this case it is again helpful to invoke the Hamiltonian of the
ferromagnetic $XY$ model:
\begin{equation}
H = -\sum_{\langle j,k \rangle} \Big(a^\dag_k a_j
+a^\dag_j a_k\Big).
\end{equation}
Here the sum is over all links in the square lattice.  The optimal
value of $K$ is the infinite-lattice limit of $(-E_0/L)$, $E_0$
being the minimum eigenvalue of this Hamiltonian.  In the thermodynamic
limit, the ground state of the $XY$ model on a square lattice
breaks the SO(2) symmetry of the problem
and picks out a preferred direction of magnetization in the
$x$-$y$ plane \cite{KLS,KK}, which can
be characterized by a single angle $\alpha$.  
But if we choose to do so (in order to simplify the interpretation
of parallel transport), we can easily generate 
a rotationally invariant state with the same 
energy---or in our context, with the same degree
of correlation---simply by averaging the ground state density matrix
$\rho(\alpha)$ over all angles: 
$\rho_0 = (1/2\pi)\int\rho(\alpha)d\alpha$.  
The ground-state energy has been
evaluated numerically \cite{HHB,ZR,SH}, and one finds that 
$K_{\max}({\mathcal A}_0) = 0.549$.  Notice that this
value is smaller than the corresponding value for a rebit
ring in the limit $n\rightarrow\infty$ (see Eqs.~(\ref{evenasymp}) 
and (\ref{oddasymp})
or Ref.~\cite{LSM}), which is
$2/\pi = 0.637$.

Given a different specification of the phases, it is not
immediately
obvious whether there exists a state that has all non-zero 
$c_{j,k}$'s---let us call such a state ``fully
connected''---and that is also consistent with the given phases.
Consider, for example, the configuration 
${\mathcal A}$ in which all the phases are zero except 
at a specific link $\langle j,k \rangle$, where the phase
is required to be $\phi$.  Can one find a fully connected 
state of the lattice 
consistent with these phases?  The following method
will work, though it is not likely to be optimal.  Start with 
the state $\rho_0$ defined in the preceding paragraph,
in which all the phases are zero.  Construct the following
sum:
\begin{equation}
\rho(j,k,\phi) = \frac{1}{3}\Big[ V_j \rho_0 V^\dag_j
+ V^\dag_k \rho_0 V_k + (V^\dag_j\otimes V_k) \rho_0 (V_j\otimes 
V^\dag_k)\Big], \label{V}
\end{equation}
where $V_j$ is the matrix
\begin{equation}
V_j = \mtx{cc}{1 & 0 \\ 0 & e^{i\xi}}
\label{singlelink}
\end{equation}
applied to particle $j$ and $V_k$ is the same
matrix applied to particle $k$.  Here $\xi$ will be 
a function of $\phi$ to be determined later.
Because the rotations in Eq.~(\ref{V}) affect only
particles $j$ and $k$, all links not involving either of these
particles will continue to have zero phase.  Moreover, any
link involving only one of the two special sites will 
likewise have its phase unchanged.  Consider, for example, the 
link $\langle j,l \rangle$ where $l\neq k$.  The value
of $c_{j,l}$ is 
\begin{equation}
c_{j,l} = \frac{1}{3}(c_0 e^{-i\xi} + c_0 + c_0 e^{i\xi}) = 
\bigg(\frac{1+2\cos(\xi)}{3}\bigg)c_0,  \label{jl}
\end{equation}
so that the phase has not been affected.  On the other
hand, the value of $c$ associated with the link $\langle j,k \rangle$
is
\begin{equation}
c_{j,k} = \bigg(\frac{2e^{-i\xi}+e^{2i\xi}}{3}\bigg)c_0,
\label{jk}
\end{equation}
which has a phase that can be made equal to 
$\phi$ by a proper choice of the value of $\xi$.

An even simpler strategy, which is surely not optimal,
shows that {\em any} phase configuration ${\mathcal A}$
can be realized in a fully connected quantum state of the lattice.
Let us imagine the two dimensions of the lattice to be
horizontal and vertical.  Start with a state in which 
each vertical column of lattice sites is in the ground state
of the ferromagnetic $XY$ model for an infinite
chain.  Call this state $|\phi_V\rangle$.  Now rotate
each of the rebits in each of these chains so as to
achieve the desired phases for the vertical links.
This can be done with no loss of correlation, because
we are simply performing local rotations.  Similarly,
consider the state $|\phi_H\rangle$ in which each
{\em horizontal} row is in the $XY$ ground state,
and rotate the rebits so as to achieve the desired
phases for the horizontal links.  Let $|\phi_V'\rangle$
and $|\phi_H'\rangle$ be the states resulting from
these rotations.  Then the mixed state
\begin{equation}
\rho = \frac{1}{2}\Big( |\phi_V'\rangle\langle\phi_V'|
+ |\phi_H'\rangle\langle\phi_H'| \Big)
\end{equation}
completely matches the phase configuration ${\mathcal A}$.
We can even compute the value of $K$ for the 
state $\rho$: it is equal to half of $K_{\max}$
for the infinite chain, independent of the 
configuration ${\mathcal A}$.  That is, $K = 1/\pi = 0.318$.
For the special configuration considered above, in which
only one link has non-zero phase, this value is smaller
than what one can achieve with the specialized method of 
Eq.~(\ref{singlelink}).  Nevertheless, the 
method we have just described does answer our 
first question: all configurations ${\mathcal A}$ can
be achieved without making any $c_{j,k}$ vanish.  
Notice also that this construction gives us
a lower bound on $K_{\max}({\mathcal A})$ for all
configurations ${\mathcal A}$: 
$K_{\max}({\mathcal A})\geq 1/\pi$.

Actually finding $K_{\max}({\mathcal A})$, even for
simple configurations ${\mathcal A}$, is probably a 
very hard problem.  If one uses strategies similar in
spirit to the one given in Eq.~(\ref{singlelink}), then
it would seem that the value of $K_{\max}({\mathcal A})$
must decrease in order to achieve non-trivial
parallel transport around a loop.  (One can see the
decrease in the values of $|c_{j,l}|$ and $|c_{j,k}|$
in Eqs. (\ref{jl}) and (\ref{jk}).)  But it is conceivable 
that a completely different strategy could do much
better; so we must leave this basic question unanswered.

\section{Lattices of qubits}

In our definition of parallel transport for rebits, developed 
in Section \ref{parallel}, we implicitly made use of the fact
that for rebits, there exists a two-particle state $\rho^{(+)}$
such that if particles $j$ and $k$ are in this state, and if 
particle $j$ is measured
and found to be in the state $|\psi\rangle$, then particle
$k$ will always be brought to the {\em same} state $|\psi\rangle$.
Mathematically, 
\begin{equation}
\omega = \frac{1}{P}\langle \psi | \rho^{(+)} 
|\psi\rangle = |\psi\rangle\langle\psi |.
\end{equation}
The state $\rho^{(+)}$ is in fact $|\Psi^{(+)}\rangle\langle\Psi^{(+)}|$,
with $|\Psi^{(+)}\rangle = (\mid\uparrow\downarrow\rangle + 
\mid\downarrow\uparrow\rangle )/\sqrt{2}$.  So for this state,
the parallel transporter is the identity and the degree of correlation
is 1.  In effect, our definition of parallel transport compares other
two-particle states to this special state; when the parallel transporter $U$ is 
not the identity, it is because one of the particles has been rotated
(and possibly distorted in other ways as well) compared to the
standard state $\rho^{(+)}$.

For qubits, there is no two-particle state with this property.  
The most closely analogous state is the singlet state
$|\Psi^{(-)}\rangle = (\mid\uparrow\downarrow\rangle - 
\mid\downarrow\uparrow\rangle )/\sqrt{2}$.  It has the property that
if a measurement on one of the particles brings it to the state
$|\psi\rangle$, the other particle will be brought to the 
{\em orthogonal} state $|\tilde{\psi}\rangle = \sigma_y|\psi^*\rangle$,
where the complex conjugation is in the standard basis.  
In defining parallel transport for qubits, we will take the singlet 
state as our standard state, for which the parallel transporter is 
defined to be the
identity.  All other states will then be compared with this one.
A general parallel transporter will be a rotation of the Bloch sphere,
that is, an element of SO(3); as in the case
of rebits, our definition will not allow us to distinguish overall
phases.  We will usually represent such a rotation as a $2\times 2$
unitary matrix, keeping in mind that the overall phase
is irrelevant.

We start again with Eq.~(\ref{map}): 
\begin{equation}
\omega^{(k)} = \frac{1}{P}\langle \psi^{(j)} | \rho^{(jk)} |\psi^{(j)}\rangle,
\end{equation}
and again let ${\mathcal M}$ be the map that 
takes $|\psi^{(j)}\rangle$ to $\omega^{(k)}$.
But now we define our generalized fidelity as
\begin{equation}
F(R) = \frac{\int \langle \tilde{\psi} | R^{\dag} {\mathcal M}(\psi) R 
| \tilde{\psi}\rangle P d\psi}
{\int P d\psi},
\label{qubitF}
\end{equation}
where $d\psi$ refers to the uniform measure over the surface
of the Bloch sphere.
Note that $F$ compares ${\mathcal M}(\psi)$ with a rotated version of 
$|\tilde{\psi}\rangle$ rather than a rotated version 
of $|\psi\rangle$.  If there is a unique rotation $R = U$ (up to an overall
phase) that
maximizes $F(R)$, then we will take this $U$ to be the parallel transporter
for the link $\langle j,k \rangle$.  Thus, if $\rho^{(jk)}$ happens
to be the singlet state, we have $U(k,j) = I$.  

Carrying out the integrals in Eq.~(\ref{qubitF}), we find that
\begin{equation}
F(R) = \frac{1}{3} + \frac{2}{3}
\langle\Psi^{(-)}|(I\otimes R^\dag)\rho^{(jk)}(I\otimes R)
|\Psi^{(-)}\rangle.
\end{equation}
Now, it is a fact that any maximally entangled state of two
qubits can be written as $(I\otimes R)|\Psi^{(-)}\rangle$.
So maximizing $F(R)$ over all rotations $R$ is the same as
finding the maximally entangled state that has the greatest
overlap with $\rho^{(jk)}$.  The quantity
\begin{equation}
f = \max_R\langle\Psi^{(-)}|(I\otimes R^\dag)\rho^{(jk)}(I\otimes R)
|\Psi^{(-)}\rangle
\end{equation}
has been called the ``fully entangled
fraction'' of $\rho^{(jk)}$ \cite{BDSW}.  It ranges from 1/4 (for the 
completely mixed state) to 1 (for a maximally entangled
state).\footnote{Even 
though the word ``entangled''
appears in the description of $f$, it is not
a proper measure of entanglement; a separable
state can have a value of $f$ greater than 1/4.}  
We will take $(4f-1)/3$, which ranges from 0 to 1,
as our measure of the degree of correlation between
particles $j$ and $k$, and define $K_q$ ($q$ for ``qubit'') 
to be the average
of this quantity over all the links of the lattice.  As
has been mentioned, the rotation $R$ that achieves this
maximum value, if it is unique, will be our parallel
transporter $U(k,j)$.\footnote{In the case of qubits, in contrast to
that of rebits, it is possible for $R$ not to be 
unique even when the correlation $(4f-1)/3$ is not 
zero.  For example, if $\rho^{(jk)}= (\mid\uparrow\downarrow\rangle
\langle\uparrow\downarrow\mid +$
\hbox{$\mid\downarrow\uparrow\rangle\langle\downarrow\uparrow
\mid)/2$}, then any rotation around
the $z$ axis maximizes $F(R)$, and yet the correlation
$(4f - 1)/3$ has the value 1/3.  However, as in the case of rebits,
this sort of ambiguity does not cause any difficulties
for the problem we are considering.}  We would like to maximize
$K_q$ for a fixed set of parallel transporters.

Our notion of parallel transport for qubits is particularly
simple if $\rho^{(jk)}$ happens to be a ``twisted Werner state,''
that is, a state of the form \cite{VW}
\begin{equation}
p (I\otimes V) |\Psi^{(-)}\rangle\langle\Psi^{(-)}|(I\otimes V^\dag) 
+ (1-p)(I/4). \label{twisted}
\end{equation}
Here $0 < p \leq 1$, $|\Psi^{(-)}\rangle$ is the singlet state,
$I$ is the $4\times 4$ identity matrix, 
and $V$ is a unitary transformation acting 
on particle $k$.  
If $\rho^{(jk)}$ is of this form, then the parallel transporter $U(k,j)$
works out, not surprisingly, to be the transformation $V$.
Moreover, the weight $p$ appearing in Eq.~(\ref{twisted}) 
is none other than our measure of correlation
$(4f-1)/3$.  Thus both the parallel transporter and the
degree of correlation are particularly easy to interpret 
in this case.  One can show that the six-rebit state of 
Eq.~(\ref{six}), reinterpreted 
as the state of a six-qubit ring, has the property that each
pair of nearest neighbors is of the twisted Werner form.

A more interesting example of a qubit ring 
exhibiting non-trivial 
parallel transport---but whose pairs are not necessarily 
of the twisted Werner
form---is given by the following state of 
six qubits.
\begin{eqnarray}
|\psi\rangle = \alpha(\mid\uparrow\uparrow\uparrow
\downarrow\downarrow\downarrow\rangle - \cdots) +
\beta[e^{-i\xi}(|\uparrow\downarrow\uparrow\downarrow
\downarrow\uparrow\rangle - \cdots) + e^{i\xi}(\mid\uparrow
\uparrow\downarrow\downarrow\uparrow\downarrow\rangle - \cdots)] \\
+ \gamma(\mid\downarrow\uparrow\downarrow\uparrow
\downarrow\uparrow\rangle - 
\mid\uparrow\downarrow\uparrow
\downarrow\uparrow\downarrow\rangle). \nonumber
\end{eqnarray}
Here each ellipsis indicates all the translations of the
given state, but with {\em alternating} signs.  For example,
the coefficient $\alpha$ multiplies
\begin{equation}
\mid\uparrow\uparrow\uparrow\downarrow\downarrow\downarrow\rangle -
\mid\uparrow\uparrow\downarrow\downarrow\downarrow\uparrow\rangle +
\mid\uparrow\downarrow\downarrow\downarrow\uparrow\uparrow\rangle -
\mid\downarrow\downarrow\downarrow\uparrow\uparrow\uparrow\rangle +
\mid\downarrow\downarrow\uparrow\uparrow\uparrow\downarrow\rangle -
\mid\downarrow\uparrow\uparrow\uparrow\downarrow\downarrow\rangle .
\end{equation}
The coefficients $\alpha$, $\beta$ and $\gamma$ are real and 
positive; their values will be specified shortly.  
Note that each pair of nearest neighbors in this state
has the same density matrix, so that each has the same degree 
of correlation and the same parallel transporter.  
Because the state $|\psi\rangle$ is an eigenstate of $S_z$
with eigenvalue zero,
the density matrix of each pair is of the form (\ref{rho}) with
all the $x_i$'s equal to zero.  (It is convenient to 
introduce a negative sign in the
off-diagonal elements since our standard state is now the
singlet.)
\begin{equation}
 \rho^{(jk)} = \mtx{cccc}{a&0&0&0\\0&b&-c&0
\\0&-c^*&b&0\\0&0&0&a}.
\end{equation}
Carrying out the trace of $|\psi\rangle\langle\psi|$ over the other particles, 
one finds that $a = 2\alpha^2 + 2\beta^2$, 
$b = \alpha^2 + 4\beta^2 + \gamma^2$, and
$c = 2\beta[(\alpha+\gamma)e^{i\xi} + \beta e^{-2i\xi}]$.
The correlation $K_q$ is $(4f-1)/3$, where
\begin{equation}
f = b + |c| = \alpha^2 + 4\beta^2 + \gamma^2
+ 2\beta \Big|(\alpha+\gamma)e^{i\xi} + \beta e^{-2i\xi}\Big|,
\end{equation}
and the parallel transporter is
\begin{equation}
U = \mtx{cc}{1 & 0 \\ 0 & e^{i\phi}},  \label{silly}
\end{equation}
where $\phi$ is the phase of $c$, that is, the phase
of $(\alpha+\gamma)e^{i\xi} + \beta e^{-2i\xi}$. The 
fact that $K_q$ depends on the matrix element $b$, while the
analogous quantity $K$ for rebits depended only on $c$, 
ultimately comes from the fact that 
in Eq.~(\ref{qubitF}) we average over the 
entire surface of the Bloch
sphere and not just over the equator. 

In the spirit of our main problem, we would like to choose
$\alpha$, $\beta$ and $\gamma$ so as to maximize
$K_q$.  Let us do this extremization for the special
case $\xi = 0$, for which the parallel transporter given
in Eq.~(\ref{silly}) is the identity.  In this case one finds that the 
optimal values are
$\alpha = (130+34\sqrt{13})^{-1/2}$, 
$\beta = (1/2)(3+\sqrt{13})\alpha$, and 
$\gamma = (4 + \sqrt{13})\alpha$,
which satisfy the normalization condition 
$6\alpha^2 + 12\beta^2 + 2\gamma^2 = 1$.\footnote{By no
accident, the state $|\psi\rangle$ with these values of the
coefficients is the ground state of the antiferromagnetic
Heisenberg model for a ring of six qubits.}  In what follows we
will assume that $\alpha$, $\beta$ and $\gamma$ have these 
values.  One finds then that for $\xi=0$,
each pair of nearest neighbors is in a Werner state, and the
correlation $K_q$ is 
$(2 + \sqrt{13})/9 = 0.623$.  Let us call this value $K_q^{(0)}$.  

How do $K_q$ and $U$ for the state $|\psi\rangle$ change
as $\xi$ departs from zero?  Let us first look at $U$.  To lowest 
order in $\xi$, the 
angle $\phi$, which is the rotation angle associated with parallel
transport across a link, is  
\begin{equation}
\phi = \hbox{arg}[(\alpha+\gamma)(1+i\xi)+\beta (1-2i\xi)]
= \bigg(\frac{\alpha+\gamma-2\beta}{\alpha+\gamma+\beta}\bigg)\xi.
\end{equation}
Note that $\alpha+\gamma - 2\beta >0$, so that this linear
contribution to $\phi$ does not vanish.  Meanwhile, the correlation
$K_q$ diminishes by an amount proportional to the square of $\xi$:
\begin{equation}
K_q = K_q^{(0)} - \bigg(\frac{12\beta^2(\alpha+\gamma)}
{\alpha+\gamma+\beta}\bigg)\xi^2.
\end{equation}
Letting $\theta = 6\phi$ be the net rotation associated
with parallel transport around the whole ring, we can
see how $K_q$ depends on $\theta$ to lowest order:
\begin{equation}
K_q = K_q^{(0)} - \bigg(\frac{\beta^2(\alpha+\gamma)
(\alpha+\gamma+\beta)}
{3(\alpha+\gamma-2\beta)^2}\bigg)\theta^2 = 0.623 - (0.369)\theta^2.
\end{equation}
Notice that the reduction in correlation is of second order
in $\theta$, whereas in the case of rebits it is of first 
order as seen in Eqs.~(\ref{evenasymp}) and (\ref{oddasymp}).
Of course we have not done the thorough 
optimization for qubits that we have done for rebit
rings, but this example indicates that there is a
significant difference between the two cases.

\section{Discussion}

We have shown, first of all, that there is a sense in which
certain quantum states exhibit non-trivial parallel transport
around a closed loop, which is to say that the nearest-neighbor
correlations exhibit a net twist as one goes around the loop.
One might regard this result as somewhat surprising, since
there is, after all, only a single quantum state for the whole
loop, and one might think that the local twists would therefore
have to cancel each other out.
We have also shown that in the case of the rebit ring, there is
a loss of nearest-neighbor correlation associated
with non-trivial parallel transport around the ring.  
For other lattices or for lattices of qubits, we do not
know whether there is such a correlation cost, though
it is certainly plausible that there would be.  

As we have seen, there is a close relationship between our rebit 
problem and
the $XY$ model, a model that has been very well studied.  
Studies of spin stiffness in this model (see, for example, Ref.~\cite{SH})
have a certain similarity with the problem we have been
considering in that in both cases one enforces a twist between
neighboring spins.  What distinguishes the present work is the fact that
we have not actually specified any 
Hamiltonian.  Though it has
been helpful for us to use an operator similar to the Hamiltonian
of the $XY$ model, we are really working with what might
be called the {\em kinematics} of quantum mechanics.
We are asking what correlation properties of quantum states
follow from certain other
correlation properties---specifically, we are asking
what one can say about the {\em strength}
of correlations given some information about the 
{\em twist} in the correlations---and this question is 
independent of any considerations of energy.

I introduced the subject by relating it to lattice gauge theory.
To what extent, then, have we found an analogy between the state of
a quantum lattice and the configuration of a lattice gauge field?
In a qualitative sense, the reduction in correlation that we have
observed in a rebit ring can be compared
to the reduction in probability that one finds in a lattice gauge 
theory.  
But for this rebit case, the 
analogy must be regarded as quite rough, because 
there is a significant lack of congruence in the details.   
In gauge theory, the initial decrease in the probability is
of {\em second order} in the net rotation angle $\theta$ associated
with transport around a plaquette.  
In the U(1) theory, for example, the function $\exp(-S)$, with 
$S$ given by 
Eq.~(\ref{action}), decreases in proportion to $\theta^2$ for small
values of $\theta$.  In contrast, in the rebit ring, the average
correlation $K$, as given by Eqs.~(\ref{answer}) and 
(\ref{oddanswer}), decreases in proportion to $\theta$ itself.
The second-order dependence is in fact important in
lattice gauge theory for taking the continuum limit.  Moreover, 
our first-order dependence makes $K_{\max}(\theta)$
a non-analytic function, since near $\theta=0$ it takes the form
$a - b|\theta |$.  So this difference is not trivial.  

On the other hand, we have just seen that the {\em qubit}
correlation as we have defined it does seem to diminish quadratically
in $\theta$, at least for a ring of six qubits.
It is interesting to ask whether in the case of a two-dimensional
or higher-dimensional lattice, the dependence of 
$K_{\max}({\mathcal A})$ is of second order in the 
rotation angles.  In the one relevant example we have considered
for a rebit lattice,
namely, the strategy given in Eq.~(\ref{singlelink}), the
value of $K$ decreases as $\phi^{2/3}$, which is
an even sharper dependence than in the rebit ring.  
But we have not explored at all fully the range of possible 
states that one might consider for these higher-dimensional
lattices.  

It is worth commenting on the fact that in the case of 
qubits, our approach makes the parallel transporters 
elements of SO(3), whereas one might have expected
SU(2).  The nature of our definition does not allow us to 
pick out a relative phase in the relation between 
neighboring qubits.
For example, we cannot distinguish between the identity operation
and a rotation by $2\pi$, even though a pure qubit state
experiencing the latter rotation picks up a phase factor
of -1.  It is conceivable that by taking into account the density 
matrix of an entire loop, in addition to the density matrices
of the neighboring pairs, one might be able to 
make sense of this distinction
as it applies to the {\em net} rotation associated with
the loop as a whole. 

There are other ways in which one might modify the 
problem we have been considering.  One could use
a different measure of correlation or entanglement.
Moreover, even if one continues to use the quantity 
$2|c|$ for rebits and the quantity $(4f-1)/3$
for qubits
as the measure of nearest-neighbor correlation, one
could combine the correlations from all the links 
in a different way.  For example, in the case of a ring
it would make some
sense to consider the {\em product} of the individual
correlations rather that the average; this measure
has the pleasing feature that it vanishes 
if any of the links in the ring is broken.

Again, our main conclusion is this: in the one example
we have worked out in detail, if there is a non-trivial twist in
the nearest-neighbor correlations, that is, a twisting
that cannot be undone by local rotations, then there is
a corresponding reduction in the maximum possible
magnitude of these correlations.
That is, in this one example at least, twisted correlations
are weaker correlations.
This conclusion follows from the structure of the 
quantum state space and is true irrespective of the 
system's Hamiltonian.  

\vspace{1cm}

\noindent{\large{\bf Acknowledgments}}

\vspace{0.5cm}

\noindent I would like to thank John Preskill, Tim Havel, and Daniel
Aalberts for a number of
helpful suggestions.  I am also
grateful for the hospitality of the Institute 
for Theoretical Physics 
in Santa Barbara, where some of this work was done.  
This research was supported in part by the National
Science Foundation under Grant No. PHY99-07949.

\end{document}